\documentclass[aps,showkeys,showpacs,superscriptaddress,onecolumn,floatfix]{revtex4}
\usepackage{graphicx}
\usepackage{amsmath}
\usepackage{amsfonts}
\usepackage{color}

\begin{document}

\title{Majority-vote model on $(3,4,6,4)$ and $(3^4,6)$ Archimedean lattices}

\author{F. W. S. Lima}
\email{wel@ufpi.br}
\affiliation{
Departamento de F\'{\i}sica, 
Universidade Federal do Piau\'{\i},\\
57072-970 Teresina, Piau\'{\i}, Brazil
}

\author{K. Malarz}
\homepage{http://home.agh.edu.pl/malarz/}
\email{malarz@agh.edu.pl}
\affiliation{
Faculty of Physics and Applied Computer Science,
AGH University of Science and Technology,\\
al. Mickiewicza 30, PL-30059 Krak\'ow, Poland
}

\date{\today}

\begin{abstract}
On Archimedean lattices, the Ising model exhibits spontaneous ordering.
Two examples of these lattices of the majority-vote model with noise are considered and studied through extensive Monte Carlo simulations.
The order/disorder phase transition is observed in this system.
The calculated values of the critical noise parameter are $q_c=0.091(2)$ and $q_c=0.134(3)$ for $(3,4,6,4)$ and $(3^4,6)$ Archimedean lattices, respectively.
The critical exponents $\beta/\nu$, $\gamma/\nu$ and $1/\nu$ for this model are 0.103(6), 1.596(54), 0.872(85) for $(3,4,6,4)$ and 0.114(3), 1.632(35), 0.98(10) for $(3^4,6)$ Archimedean lattices.
These results differs from the usual Ising model results and the majority-vote model on so-far studied regular lattices or complex networks.
The effective dimensionalities of the system [$D_{\text{eff}}(3,4,6,4)=1.802(55)$ and $D_{\text{eff}}(3^4,6)=1.860(34)$] for these networks are reasonably close to the embedding dimension two. 
\end{abstract}



\maketitle

\section{Introduction}
 
The majority-vote model (MVM) \cite{MVM-SL} defined on regular lattices shows second-order phase transition with critical exponents $\beta$, $\gamma$, $\nu$ --- which characterize the system in the vicinity of the phase transition --- identical \cite{MVM-SL,MVM-regular,MVM-MFA} with those of equilibrium Ising model \cite{ising,critical}.

On the other hand MVM on the complex networks exhibit different behavior \cite{MVM-SW,MVM-ER,MVM-VD,MVM-AB}.
Campos {\it et al.} investigated MVM on small-world network \cite{MVM-SW}.
This network was constructed using the square lattice (SL) by the rewiring procedure.
Campos {\it et al.} found that the critical exponents $\gamma/\nu$ and $\beta/\nu$ are different from these of the Ising model \cite{critical} and depend on the rewiring probability.
Pereira {\it et al.} \cite{MVM-ER} studied MVM on Erd{\H o}s--R\'enyi's (ER) classical random graphs \cite{ER}, and Lima {\it et al.}  \cite{MVM-VD} also studied this model on random Voronoy--Delaunay \cite{VD} lattice with periodic boundary conditions.
Very recently Lima \cite{MVM-AB} studied the MVM on directed Albert--Barab\'asi (AB) network \cite{AB} and contrary to the Ising model \cite{ising} on these networks \cite{alex}, the order/disorder phase transition is observed in this system.
The calculated $\beta/\nu$ and $\gamma/\nu$ exponents are different from those for the Ising model \cite{critical} and depend on the mean value of connectivity $z$ of AB network.
The latter was observed also for ER random graph \cite{MVM-ER}.

The results obtained by these authors \cite{MVM-SW,MVM-ER,MVM-VD,MVM-AB} show that the MVM on various complex topologies belongs to different universality classes.
Moreover, contrary to MVM on regular lattices \cite{MVM-SL,MVM-regular}, the obtained critical exponents are different from those of the universality class to which the equilibrium Ising model belongs \cite{critical}.

\begin{figure*}[!hbt]
\begin{center}
\includegraphics[clip, viewport=50 64 641 395, width=.35\textwidth]{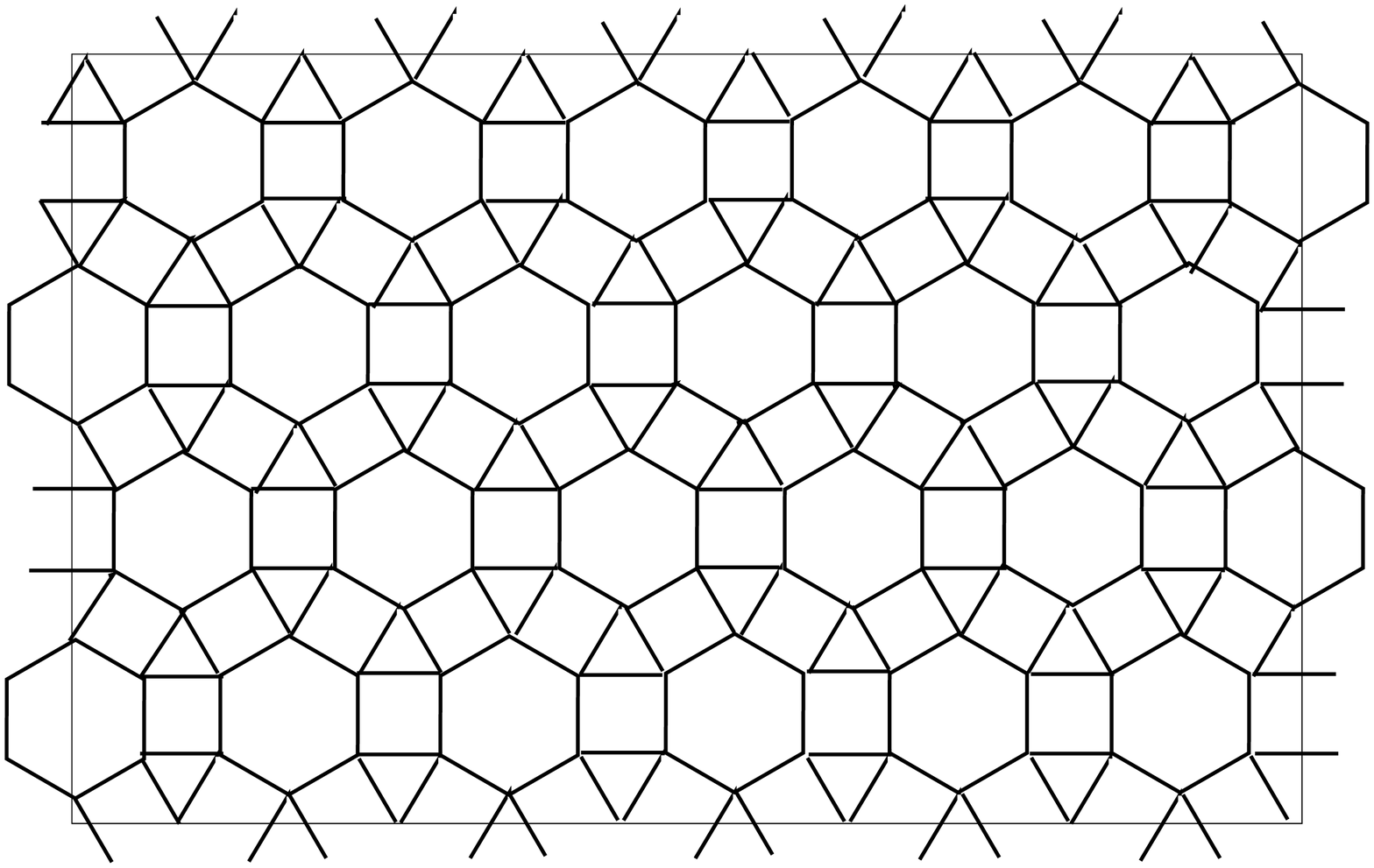}
\hspace{5mm}
\includegraphics[clip, viewport=50 64 641 395, width=.35\textwidth]{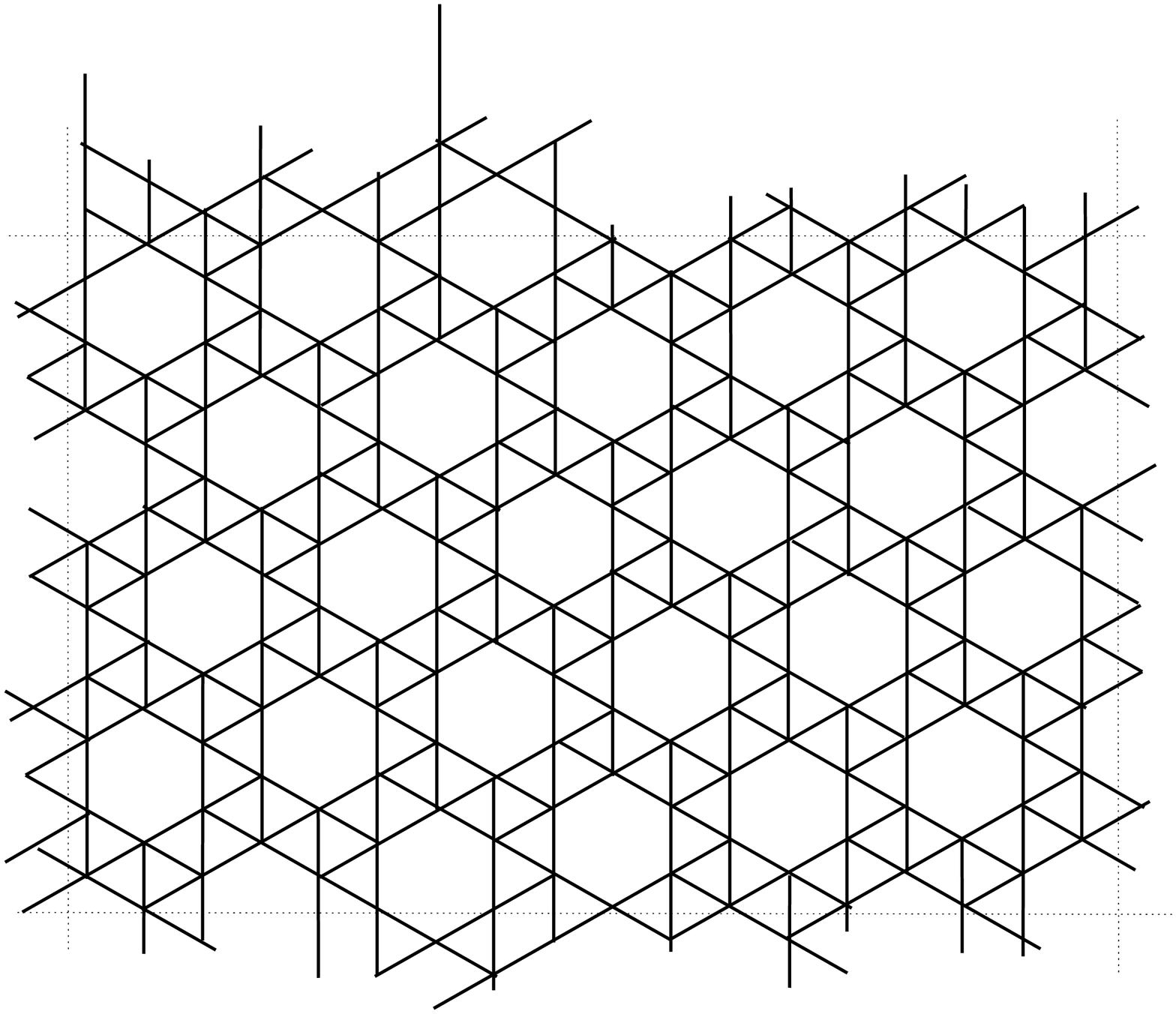}
\end{center}
\caption{\label{fig-lattice} $(3,4,6,4)$ (left) and  $(3^4,6)$ (right) AL.}
\end{figure*}

In this paper we study the MVM on two Archimedean lattices (AL), namely on $(3,4,6,4)$ and $(3^4,6)$.
The topologies of $(3,4,6,4)$ and $(3^4,6)$ AL are presented in Fig. \ref{fig-lattice}.
The AL are vertex transitive graphs that can be embedded in a plane such that every face is a regular polygon.
Kepler showed that there are exactly eleven such graphs.
The AL are labeled according to the sizes of faces incident to a given vertex.
The face sizes are sorted, starting from the face for which the list is the smallest in lexicographical order.
In this way, the square lattice gets the name $(4, 4, 4, 4)$, abbreviated to $(4^4)$, honeycomb is called $(6^3)$ and Kagom\'e is $(3, 6, 3, 6)$.
Critical properties of these lattices were investigated in terms of site percolation \cite{percol-ds} in Ref. \cite{percolation}.
Topologies of all eleven AL are given there as well.
Very recently, the Ising model on those AL was investigated in Ref. \cite{zborek}. 
 
\begin{figure*}[!hbt]
\begin{center}
\includegraphics[angle=-90,scale=.43]{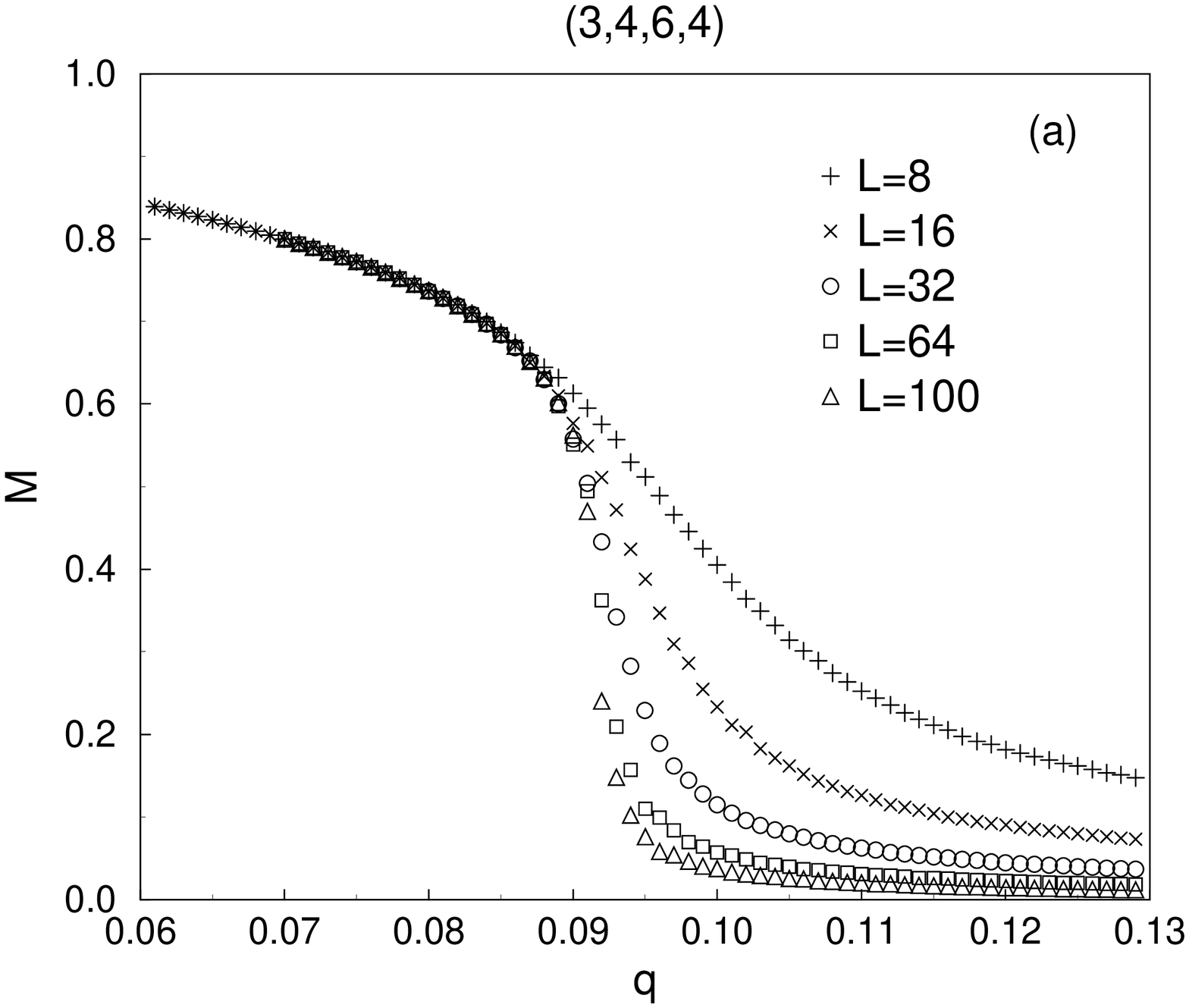}
\includegraphics[angle=-90,scale=.43]{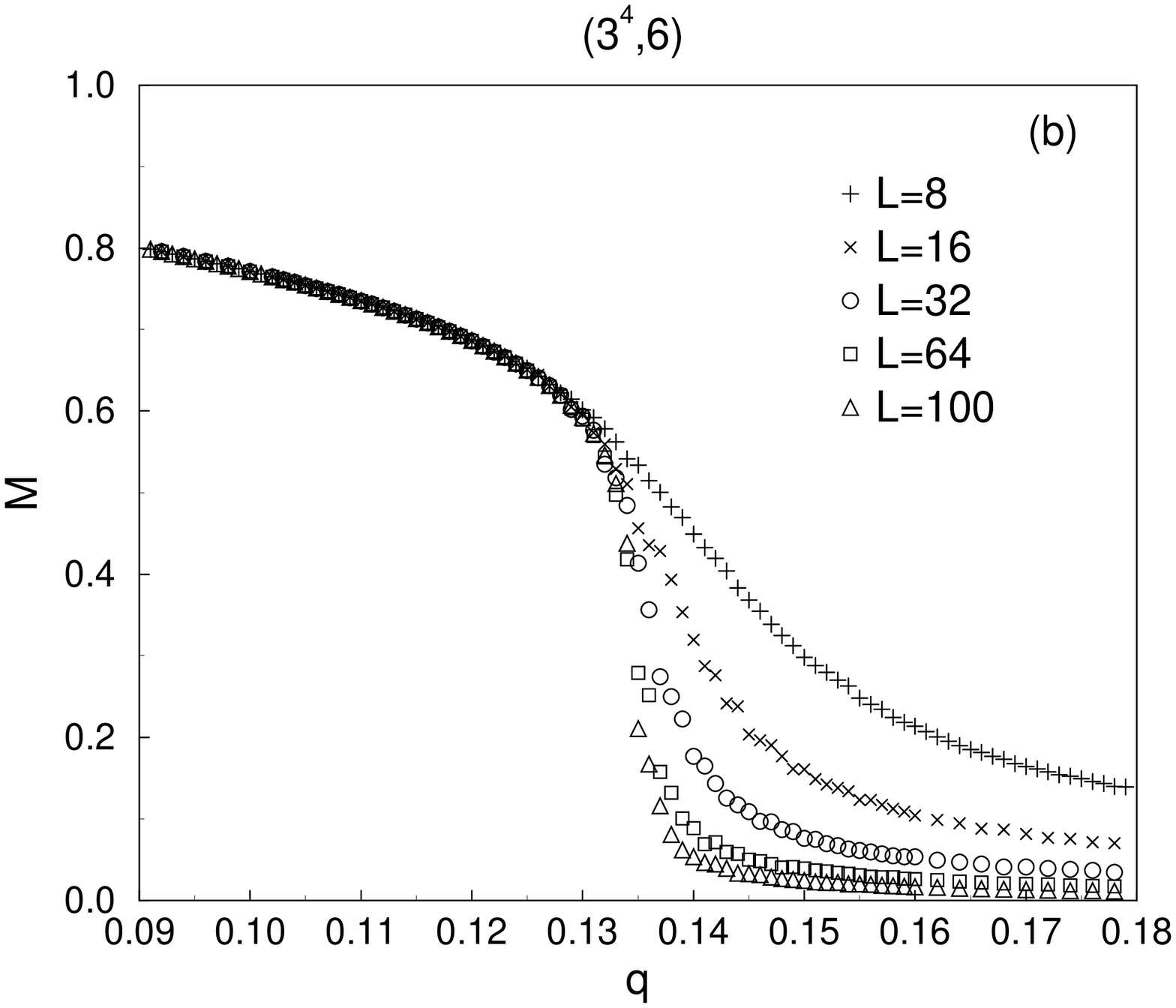}
\end{center}
\caption{\label{fig-M} The magnetization $M$ as a function of the noise parameter $q$, for $L=8$, 16, 32, 64 and 100 sites for (a) $(3,4,6,4)$ and (b) $(3^4,6)$ AL.}
\end{figure*}

Our main goal is to check the hypothesis of Grinstein {\it et al.} \cite{grinstein} --- i.e., that non-equilibrium stochastic spin systems with up-down symmetry fall in the universality class of the equilibrium Ising model --- for systems in-between ordinary, regular lattices (like square one \cite{MVM-SL}) and complex spin systems (like decorated with spin variables classical random graphs \cite{MVM-ER} or scale-free networks \cite{MVM-AB}).

With extensive Monte Carlo simulation we show that MVM on $(3,4,6,4)$ and $(3^4,6)$ {\em regular} AL exhibits second-order phase transition with effective dimensionality $D_{\text{eff}}\approx 1.8$ and has critical exponents that {\em do not} fall into universality class of the equilibrium Ising model (although for $(3^4,6)$ lattice they are quite close, yet lower).

\section{Model and simulation}

We consider the MVM \cite{MVM-SL} defined by a set of ``voters'' or spin variables $\sigma$ taking the values $+1$ or $-1$, situated on every node of the $(3,4,6,4)$ and $(3^4,6)$ AL with $N=6L^2$ sites.
The evolution is governed by single spin-flip like dynamics with a probability $w_i$ of $i$-th spin flip given by
\begin{equation}
w_i=\frac{1}{2}\left[ 1-(1-2q)\sigma_{i}\cdot\text{sign}\left(\sum_{j=1}^z\sigma_j\right)\right],
\end{equation}
where the sum runs over the number $z$ ($z(3,4,6,4)=4$ or $z(3^4,6)=5$) of nearest neighbors of $i$-th spin.
The control parameter $0\le q\le 1$ plays the role of the temperature in equilibrium systems and measures
the probability of parallel aligning to the majority of neighbors.
It means, that given spin $i$ adopts the majority sign of its neighbors with probability $q$ and the minority sign with probability $(1-q)$ \cite{MVM-SL,MVM-ER,MVM-VD,MVM-AB}.

To study the critical behavior of the model we define the variable
$m\equiv\sum_{i=1}^{N}\sigma_{i}/N$. In particular, we are interested in the
magnetization $M$, susceptibility $\chi$ and the reduced fourth-order cumulant $U$
\begin{subequations}
\label{eq-def}
\begin{equation}
M(q)\equiv \langle|m|\rangle,
\end{equation}
\begin{equation}
\chi(q)\equiv N\left(\langle m^2\rangle-\langle m \rangle^2\right),
\end{equation}
\begin{equation}
U(q)\equiv 1-\langle m^{4}\rangle/\left( 3\langle m^2 \rangle^2 \right),
\end{equation}
\end{subequations}
where $\langle\cdots\rangle$ stands for a thermodynamics average.
The results are averaged over the $N_{\text{run}}$ independent simulations.

These quantities are functions of the noise parameter $q$ and obey the finite-size
scaling relations
\begin{subequations}
\label{eq-scal}
\begin{equation}
\label{eq-scal-M}
M=L^{-\beta/\nu}f_m(x),
\end{equation}
\begin{equation}
\label{eq-scal-chi}
\chi=L^{\gamma/\nu}f_\chi(x),
\end{equation}
\begin{equation}
\label{eq-scal-dUdq}
\frac{dU}{dq}=L^{1/\nu}f_U(x),
\end{equation}
where $\nu$, $\beta$, and $\gamma$ are the usual critical 
exponents, $f_{i}(x)$ are the finite size scaling functions with
\begin{equation}
\label{eq-scal-x}
x=(q-q_c)L^{1/\nu}
\end{equation}
\end{subequations}
being the scaling variable.
Therefore, from the size dependence of $M$ and $\chi$
we obtained the exponents $\beta/\nu$ and $\gamma/\nu$, respectively.
The maximum value of susceptibility also scales as $L^{\gamma/\nu}$. Moreover, the
value of $q^*$ for which $\chi$ has a maximum is expected to scale with the system size as
\begin{equation}
\label{eq-q-max}
q^*=q_c+bL^{-1/\nu} \text{ with } b\approx 1.
\end{equation}
Therefore, the  relations \eqref{eq-scal-dUdq} and \eqref{eq-q-max} may be used to get the exponent $1/\nu$.
We evaluate also  the effective dimensionality, $D_{\text{eff}}$, from the hyperscaling hypothesis
\begin{equation}
2\beta/\nu+\gamma/\nu=D_{\text{eff}}.
\label{eq-Deff}
\end{equation}

\begin{figure*}[!hbt]
\begin{center}
\includegraphics[angle=-90,scale=.43]{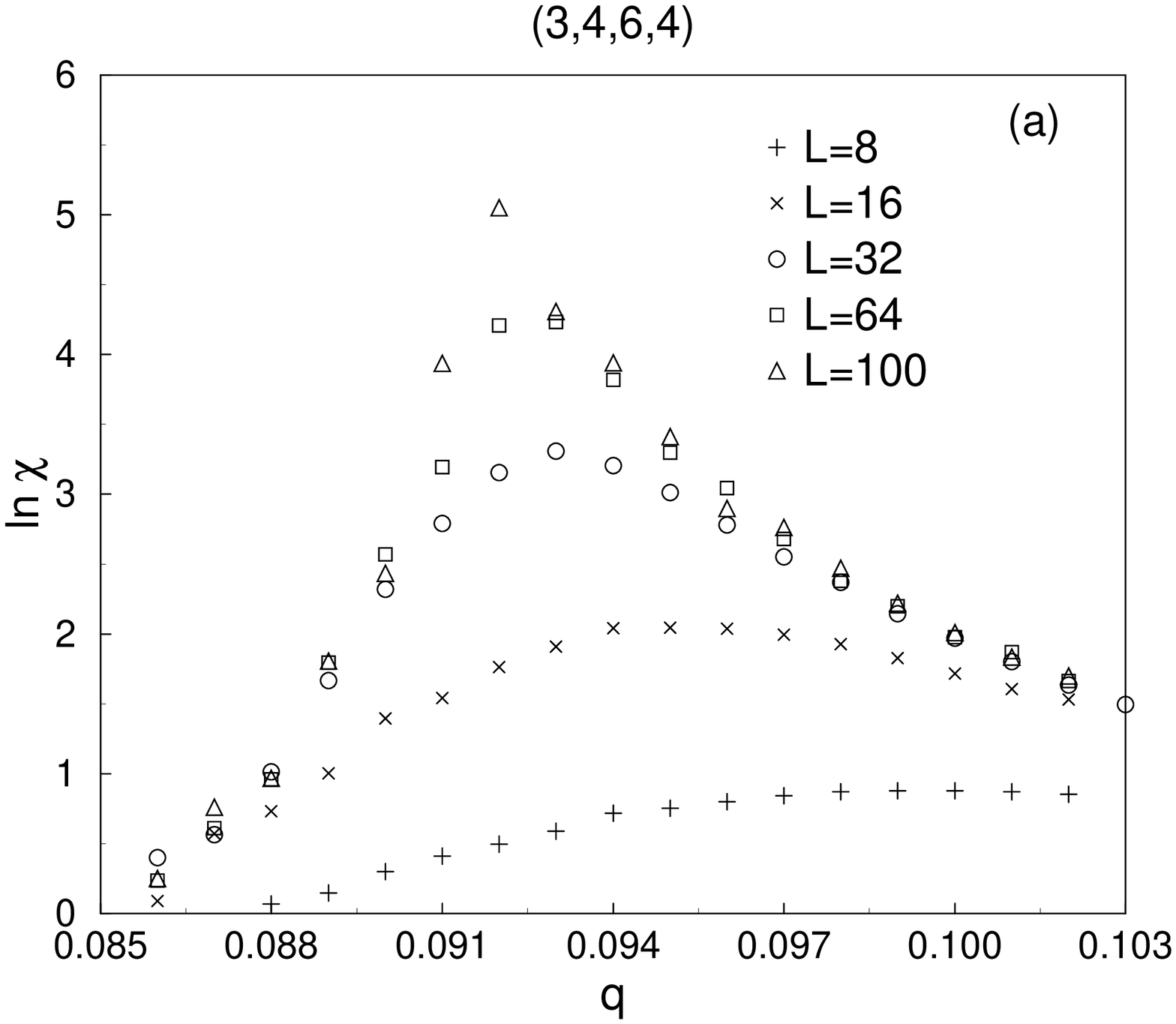}
\includegraphics[angle=-90,scale=.43]{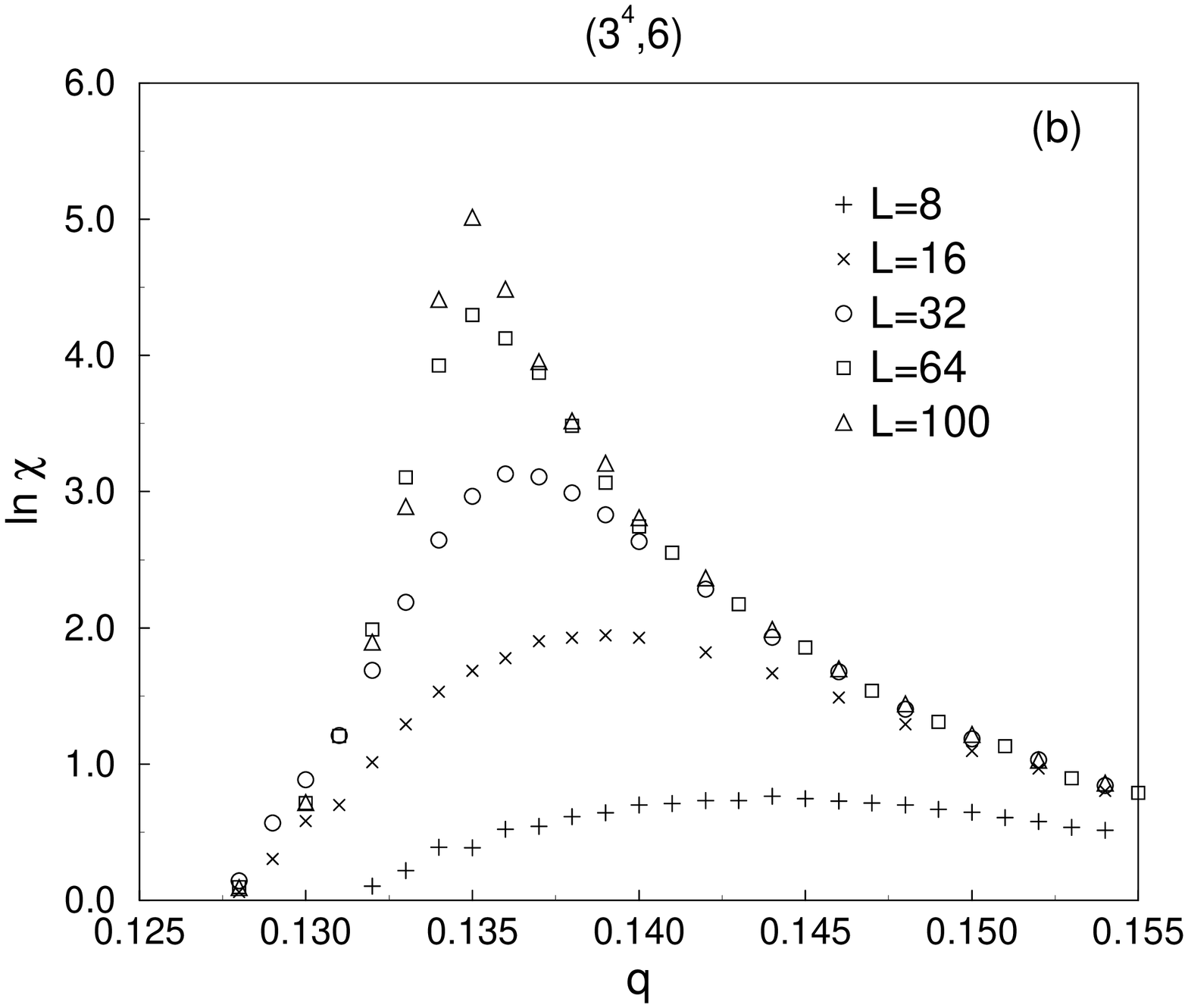}
\end{center}
\caption{\label{fig-chi} Susceptibility versus $q$ for (a) $(3,4,6,4)$ and (b) $(3^4,6)$ AL.}
\end{figure*}

We performed Monte Carlo simulation on the $(3,4,6,4)$ and $(3^4,6)$ AL with various systems of size $N=384$, $1536$, $6144$, $24576$ and $60000$.
It takes $2\times 10^5$ Monte Carlo steps (MCS) to make the system reach the steady state, and then the time averages are estimated over the next $2\times 10^5$ MCS.
One MCS is accomplished after all the $N$ spins are investigated whether they flip or not.
We carried out $N_{\text{run}}=20$ to $50$ independent simulation runs for each lattice and for given set of parameters $(q,N)$.

\section{Results and Discussion}

In Fig. \ref{fig-M} we show the dependence of the magnetization $M$ on the noise parameter $q$, obtained from simulations on $(3,4,6,4)$ and $(3^4,6)$ AL with $N$ ranging from 384 to 60000 sites.
The shape of $M(q)$ curve, for a given value of $N$, suggests the presents of the second-order phase transition in the system.
The phase transition occurs at the value of the critical noise parameter $q_c$.

\begin{figure*}[!hbt]
\begin{center}
\includegraphics[angle=-90,scale=.44]{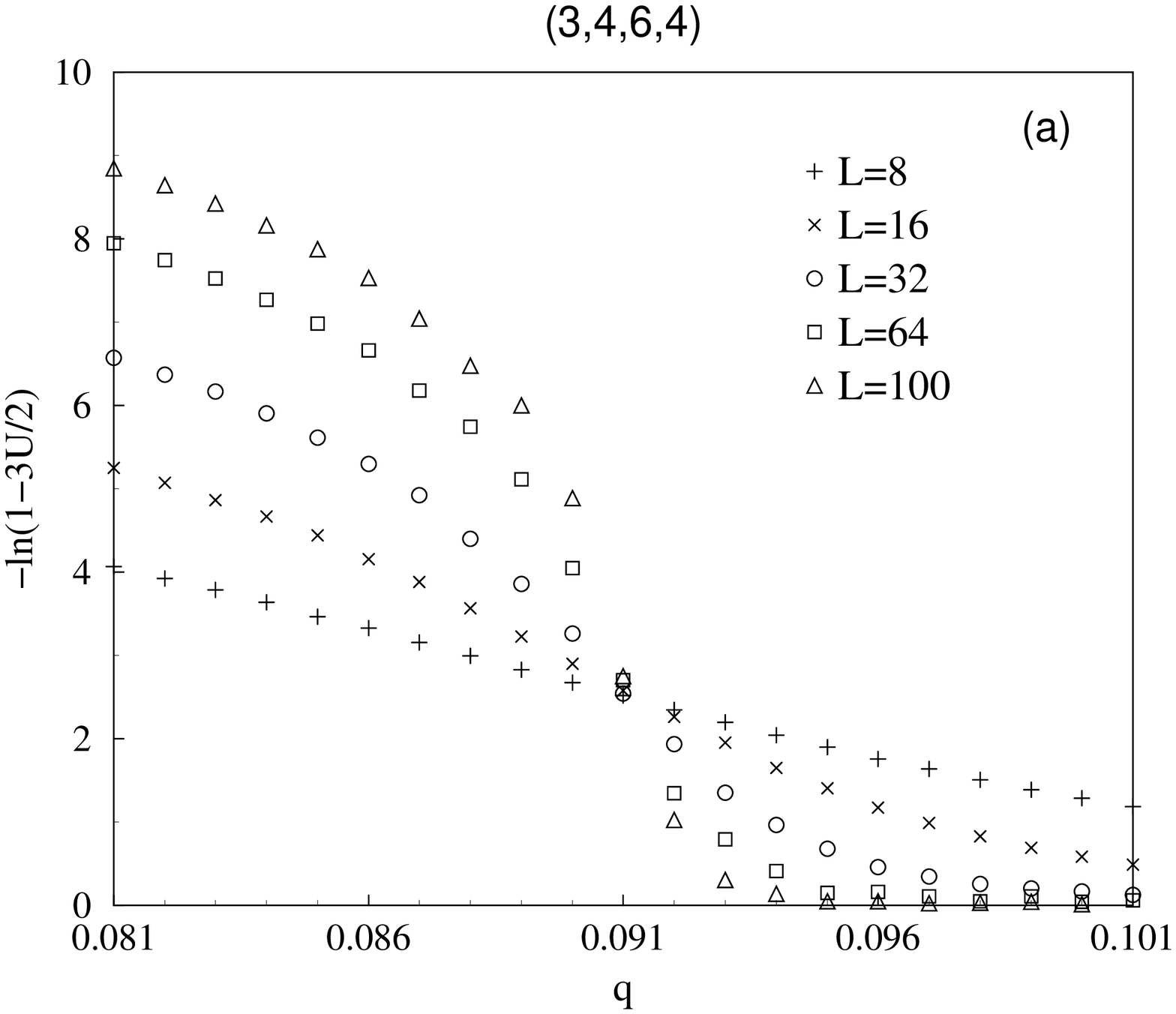}
\includegraphics[angle=-90,scale=.44]{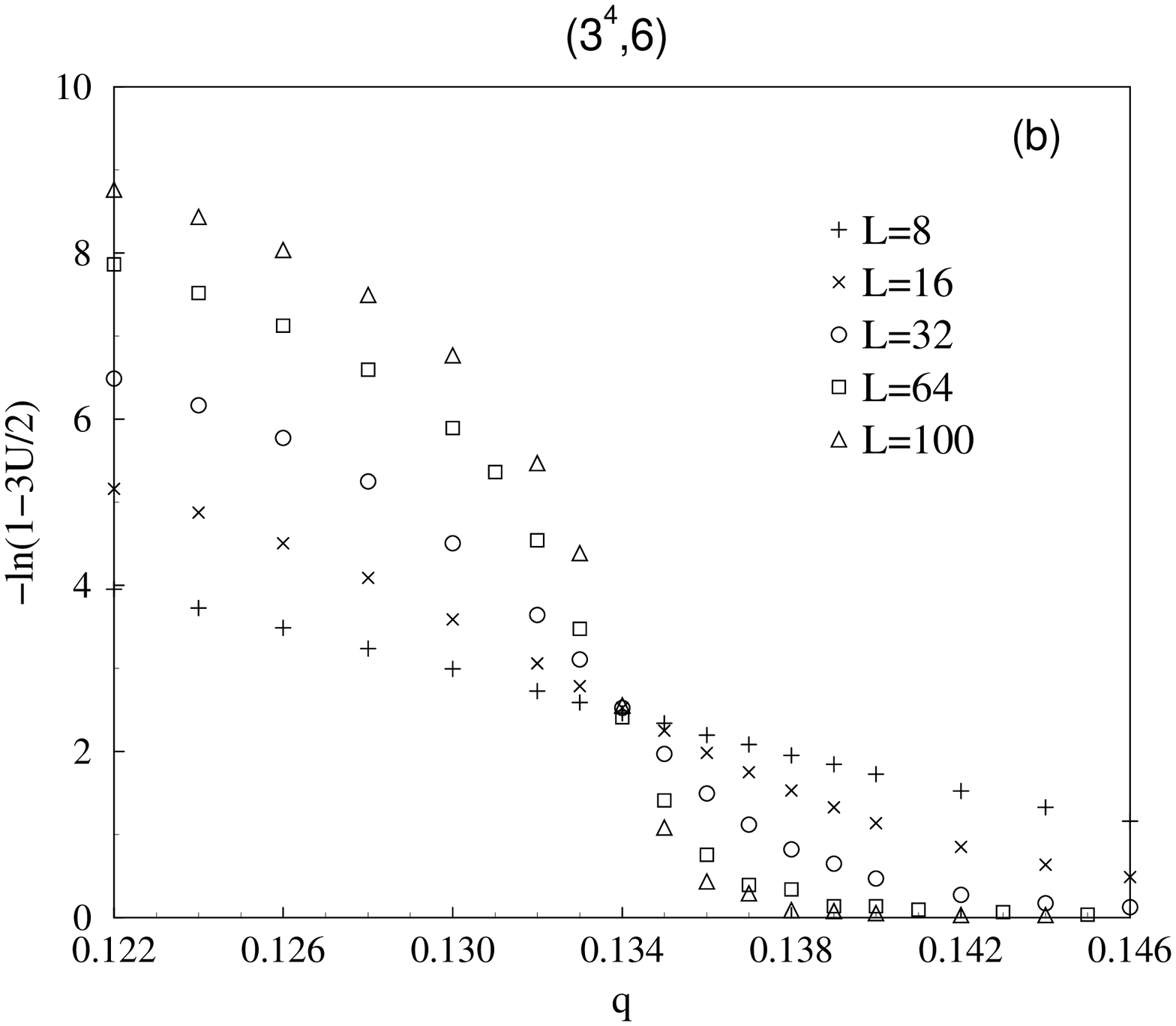}
\end{center}
\caption{\label{fig-U} The reduced Binder's fourth-order cumulant $U$ as a function of $q$ for (a) $(3,4,6,4)$ and (b) $(3^{4},6)$ AL.}
\end{figure*}

In Fig. \ref{fig-chi} the corresponding behavior of the susceptibility $\chi$ is presented.
In Fig. \ref{fig-U} we plot the Binder's fourth-order cumulant $U$ for different values of the system size $N$.
The critical noise parameter $q_c$ is estimated as the point where the curves for different system sizes $N$ intercept each other \cite{binder}.
>From Fig. \ref{fig-U} we obtain $q_c=0.091(2)$ and $q_c=0.134(3)$ for $(3,4,6,4)$ and $(3^4,6)$ AL, respectively.

\begin{figure}[!hbt]
\begin{center}
\includegraphics[angle=-90,scale=.45]{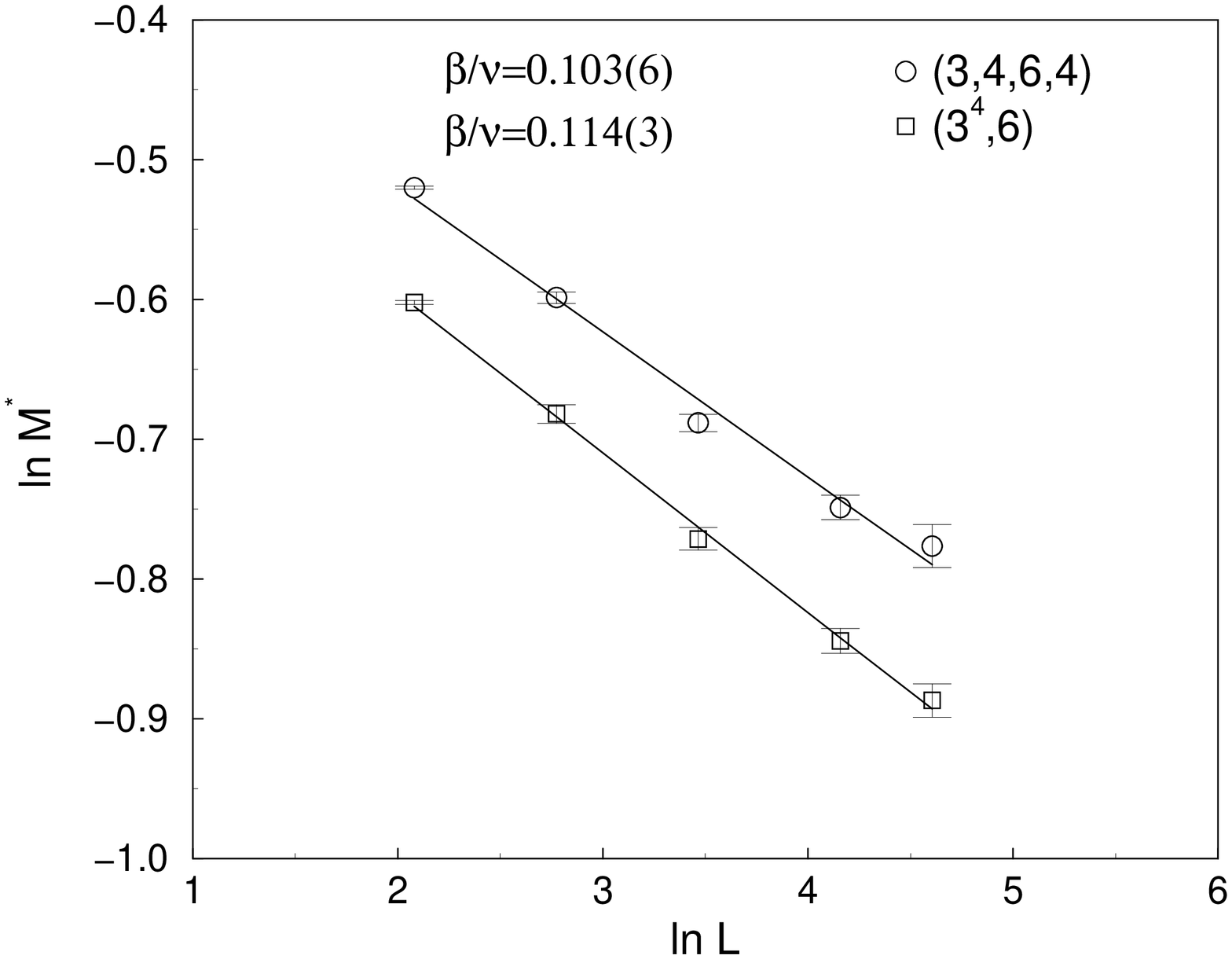}
\end{center}
\caption{\label{fig-M-N} Plot of $\ln M^*$ vs. $\ln L$ for $(3,4,6,4)$ and $(3^4,6)$ AL.}
\end{figure}

\begin{figure*}[!hbt]
\begin{center}
\includegraphics[angle=-90,scale=.45]{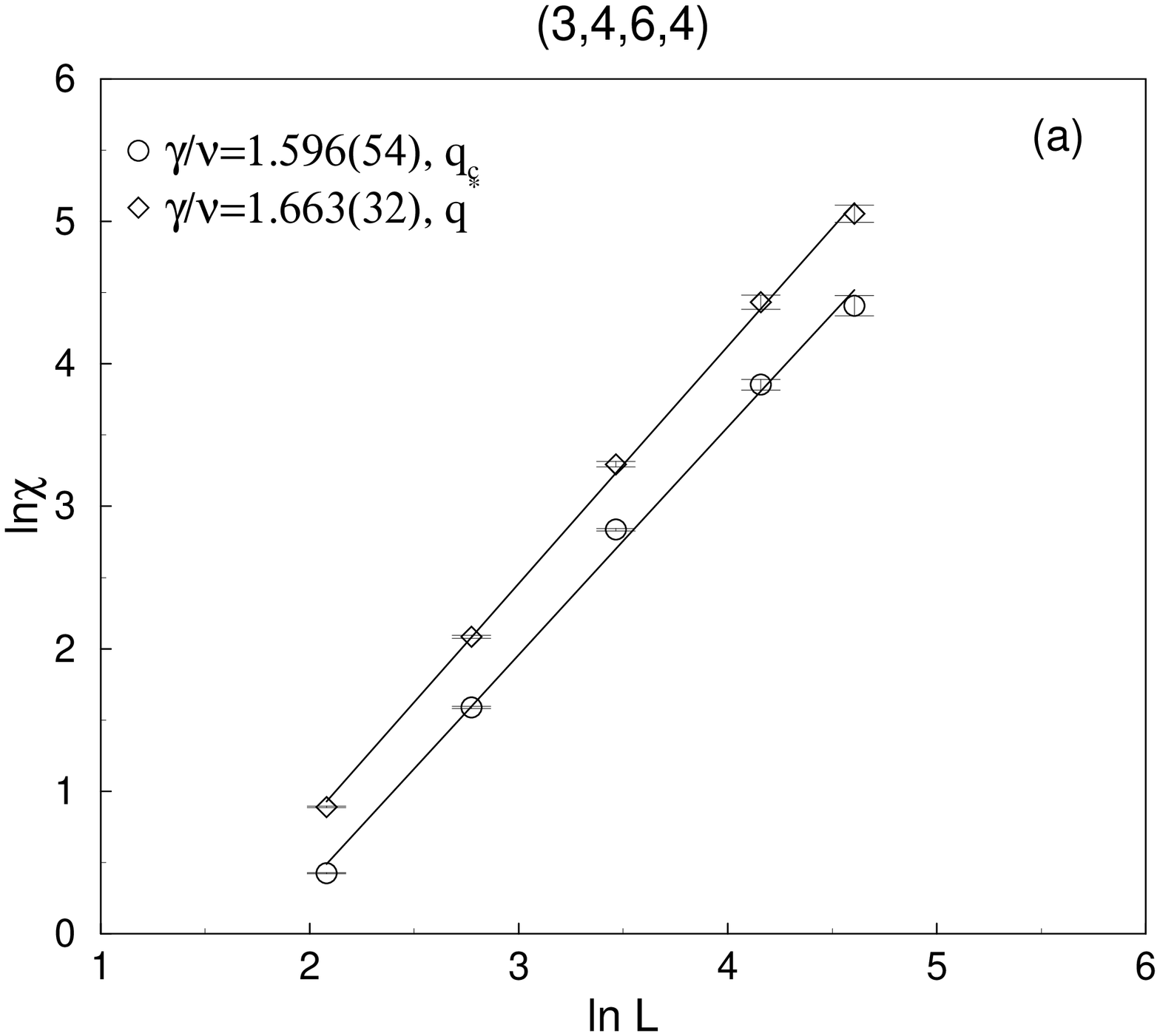}
\includegraphics[angle=-90,scale=.45]{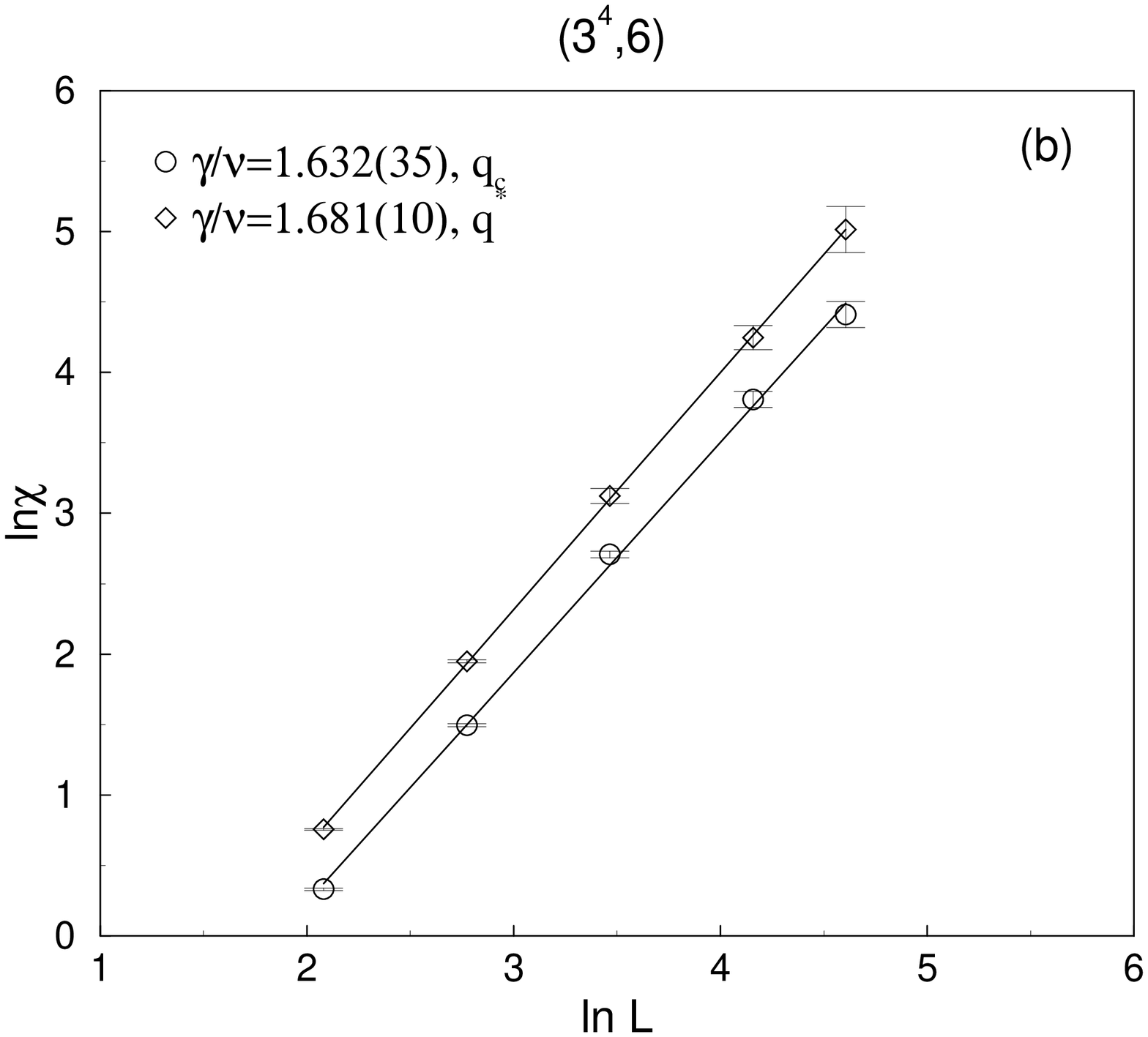}
\end{center}
\caption{\label{fig-chi-N} Critical behavior of the susceptibility $\chi(N)$ at $q=q_c$ (circle) and $q=q^*$ (diamond) for (a) $(3,4,6,4)$ and (b) $(3^4,6)$ AL.}
\end{figure*}

In Fig. \ref{fig-M-N} we plot the dependence of the magnetization $M^*=M(q_c)$ vs. the linear system size $L$.
The slopes of curves correspond to the exponent ratio $\beta/\nu$ according to Eq. \eqref{eq-scal-M}.
The obtained exponents are $\beta/\nu=0.103(6)$ and $0.114(3)$ respectively for $(3,4,6,4)$ and $(3^4,6)$ AL.
The exponents ratio $\gamma/\nu$ are obtained from the slopes of the straight lines with $\gamma/\nu=1.596(54)$ for $(3,4,6,4)$ and $\gamma/\nu=1.632(35)$ for $(3^4,6)$, as presented in Fig. \ref{fig-chi-N}.

\begin{figure}[!hbt]
\begin{center}
\includegraphics[angle=-90,scale=.45]{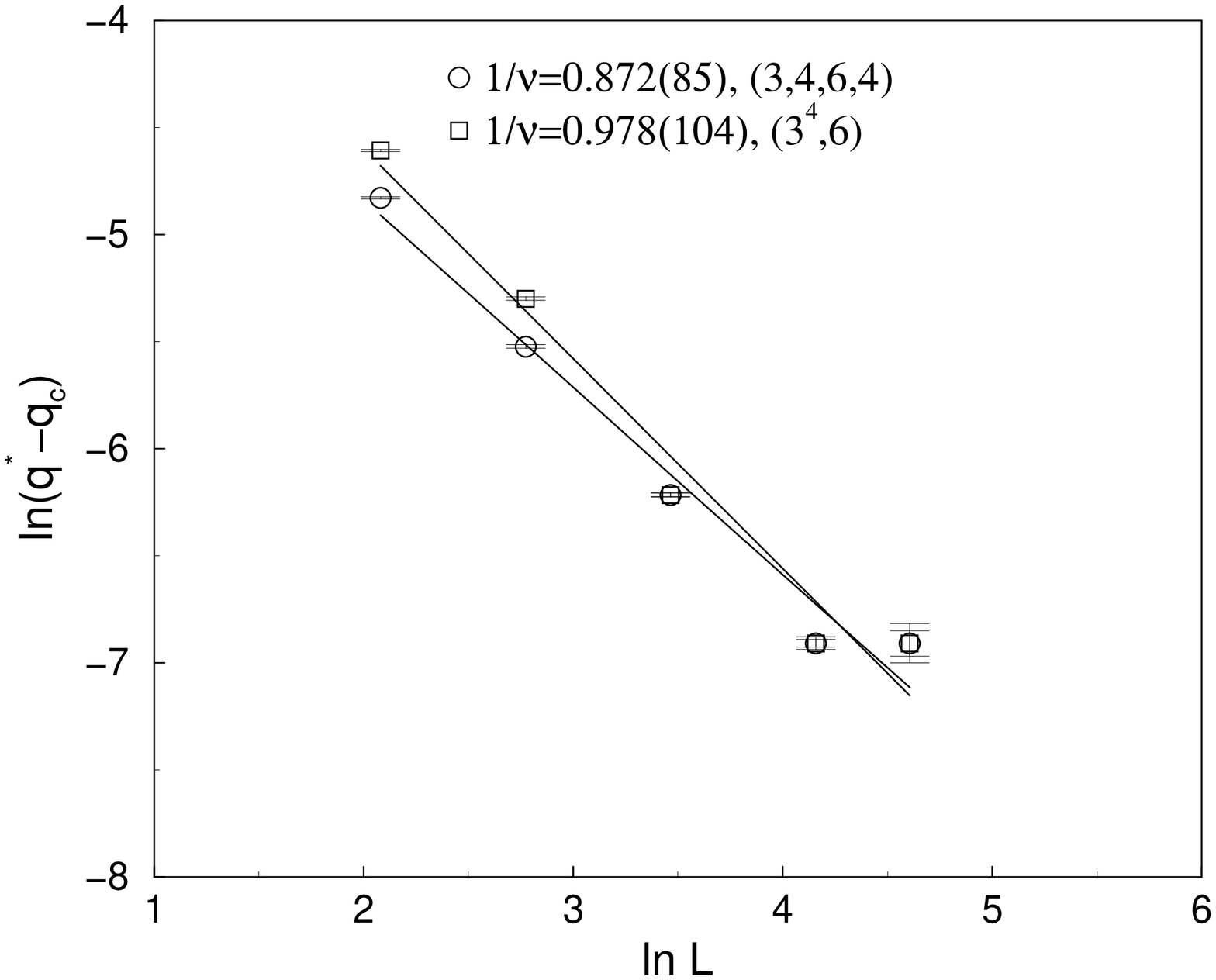}
\end{center}
\caption{\label{fig-eq-4} The exponents $1/\nu$ obtained from the relation \eqref{eq-q-max} for $(3,4,6,4)$ (circle) and  $(3^4,6)$ (square) AL.}
\end{figure}

To obtain the critical exponent $1/\nu$, we used the scaling relation \eqref{eq-q-max}.
The calculated  values of the exponents $1/\nu$ are 0.872(85) for $(3,4,6,4)$ (circle) and $1/\nu= 0.98(10)$ for $(3^4,6)$ (square) (see Fig. \ref{fig-eq-4}).
Eq. \eqref{eq-Deff} yields effective dimensionality of systems $D_{\text{eff}}=1.802(55)$ for $(3,4,6,4)$ and $D_{\text{eff}}=1.860(34)$ for $(3^4,6)$.
The MVM on those two AL has the effective dimensionality close to two contrary to ER classical random graphs ($0.99\le D_{\text{eff}}\le 1.02$) \cite{MVM-ER} or directed AB networks ($0.998\le D_{\text{eff}}\le 1.018$) \cite{MVM-AB} with roughly the same nodes connectivity ($\bar z=4$) as for $(3,4,6,4)$ and $(34,6)$ AL.

The results of simulations together with data for ER and  AB networks with mean connectivity $\bar z=4$ are collected in Tab. \ref{tab}.
 
\begin{table*}
\caption{\label{tab} Critical parameter, exponents and effective dimension for MVM model on $(3,4,6,4)$ and $(3^4,6)$ AL.
For completeness we cite data for scale-free AB and random ER graphs with an average connectivity $\bar z=4$.}
\begin{ruledtabular}
\begin{tabular}{r lllll}
 & SL \cite{MVM-SL} & AB \cite{MVM-AB} & ER \cite{MVM-ER} & (3,4,6,4) & $(3^4,6)$ \\
\hline
$q_c$
                 & 0.075(10) & 0.431(3) & 0.181(1) & 0.091(2)  & 0.134(3)  \\
$\beta/\nu$
                 & 0.125(5)  & 0.447(1) & 0.242(6) & 0.103(6)  & 0.114(3)  \\
$\gamma/\nu$\footnote{obtained using $\chi(N)$ at $q=q_c$}     
                 & 1.73      & 0.104(2) & 0.54(1)  & 1.596(54) & 1.632(35) \\
$\gamma/\nu$\footnote{obtained using $\chi(N)$ at $q=q^*$}
                 & 1.70      & 0.888(9) & 0.515(6) & 1.663(32) & 1.681(10) \\
$1/\nu$
                 & 1/0.99(5) & ---      & 0.59(7)  & 0.872(85) & 0.98(10)  \\
$D_{\text{eff}}$\footnote{obtained using ratio $\gamma/\nu$ given by dependence $\chi(N)$ at $q=q_c$}
                 & ---       & 0.998(3) & 1.02(2)  & 1.802(55) & 1.860(34) \\
\end{tabular}
\end{ruledtabular}
\end{table*}

\section{Conclusion}
 
We presented a very simple non-equilibrium MVM on $(3,4,6,4)$ and $(3^4,6)$ AL.
On these lattices, the MVM shows a second-order phase transition.
Our Monte Carlo simulations demonstrate that the effective dimensionality $D_{\text{eff}}$ is close to two, i.e. that hyperscaling may be valid.

Finally, we remark that the critical exponents $\gamma/\nu$, $\beta/\nu$ and $1/\nu$ for MVM on {\em regular} $(3,4,6,4)$ AL are {\em different} from the Ising model \cite{critical} and {\em differ} from those for so-far studied regular lattices \cite{MVM-SL,MVM-regular} and for the ER classical random graphs \cite{MVM-ER} and for the directed AB network \cite{MVM-AB}. 
However, in the latter cases \cite{MVM-ER,MVM-AB} the scaling relations \eqref{eq-scal} must involve the number of sites $N$ instead of linear system size $L$ as these networks in natural way do not posses such characteristic which allow for $N\propto L^d$ $(d\in\mathbb{Z})$ dependence \footnote{The linear dimension of such networks, i.e. its diameter --- defined as an average node-to-node distance --- grows usually logarithmically with the system size \cite{el-sw}.}.
For $(3^4,6)$ AL, the critical exponents are much closer to those known analytically for square lattice Ising model, i.e. $\beta=1/8=0.125$, $\gamma=7/4=1.75$ and $\nu=1$, but except for $\nu$ they differ for more than three numerically estimated uncertainties.
As those uncertinities were estimated basing only on statistical analysis we cannot excluding that critical exponents may reach SL Ising exponents asymptotically in the thermodynamic limit. 

\begin{acknowledgments}
FWSL acknowledges Ana Maria de Seixas Pereira for her help in the support the system SGI Altix 1350 the computational park CENAPAD, UNICAMP-USP, SP-BRASIL and also the agency FAPEPI for the financial support.
The machine time on HP Integrity Superdome in ACK\---CY\-F\-RO\-NET\---AGH is financed by the Polish Ministry of Education and Science under grant No. MNiI/\-HP\_I\_SD/\-AGH/047/\-2004.
\end{acknowledgments}
 

\end{document}